\documentclass[namedreferences]{solarphysics}
\usepackage[optionalrh,solaenum]{spr-sola-addons} % For Solar Physics 
\usepackage{graphicx}                    % For eps figures, newer & more powerfull
\usepackage{color}                       % For color text: \color command
\usepackage{url}                         % For breaking URLs easily trough lines
\usepackage{lscape}						 % landscape package, for the table with results
\usepackage{multirow}					 % multirow cells in tables
\usepackage{hyperref}

\newcommand{\etal}{{\it et al.}}
\newcommand{\insitu}{{\it in-situ}}

\newcommand{\eg}{{\it e.g.}}
\newcommand{\ie}{{\it i.e.}}

\newcommand{\f}[2]{ \frac{#1}{#2} }

\begin{document}

\begin{article}

\begin{opening}

\title{Three-Dimensional Evolution of Flux-Rope CMEs and its Relation to the Local Orientation of the Heliospheric Current Sheet}

%%%%%%%%%%%%%%%%%%%%%%%%%%%%%%%%%%%%%%%%%%%%%%%%%%%
%% Authors Names
%
\author{A.~\surname{Isavnin}$^{1}$\sep
		A.~\surname{Vourlidas}$^{2}$\sep
        E.K.J.~\surname{Kilpua}$^{1}$
       }

%%%%%%%%%%%%%%%%%%%%%%%%%%%%%%%%%%%%%%%%%%%%%%%%%%%
%% Runningheads
%
\runningauthor{A. Isavnin {\etal}}
\runningtitle{3D Evolution of Flux-Rope CMEs in Relation to the Local Orientation of the HCS}

%%%%%%%%%%%%%%%%%%%%%%%%%%%%%%%%%%%%%%%%%%%%%%%%%%%
%% Affilations 
%
  \institute{$^{1}$ Department of Physics, University of Helsinki, P.O.Box 64, FI-00014, Finland, email: \url{Alexey.Isavnin@helsinki.fi}\\
  			 $^{2}$ Space Science Division, Naval Research Laboratory, Washington DC, USA
             }

%%%%%%%%%%%%%%%%%%%%%%%%%%%%%%%%%%%%%%%%%%%%%%%%%%%
%%% Abstract 
\begin{abstract}
Flux ropes ejected from the Sun may change their geometrical orientation during their evolution, which directly affects their geoeffectiveness. Therefore, it is crucial to understand how solar flux ropes evolve in the heliosphere to improve our space-weather forecasting tools. In this article we present a follow-up study of the concepts described by \inlinecite{Isavnin2013a}. We analyze 14 coronal mass ejections (CMEs), with clear flux rope signatures, observed during the decay of Solar Cycle 23 and rise of Solar Cycle 24. First, we estimate initial orientations of the flux ropes at the origin using extreme ultraviolet observations of post-eruption arcades and/or eruptive prominences. Then we reconstruct multiviewpoint coronagraph observations of the CMEs from $\approx$~2 to 30~$\mathrm{R}_\odot$ with a three-dimensional geometric representation of a flux rope to determine their geometrical parameters. Finally, we propagate the flux ropes from $\approx$~30~$\mathrm{R}_\odot$ to 1~AU through MHD-simulated background solar wind while using {\insitu} measurements at 1~AU of the associated magnetic cloud as a constraint for the propagation technique. This methodology allows us to estimate the flux-rope orientation all the way from the Sun to 1~AU. We find that while the flux-ropes' deflection occurs predominantly below 30~$\mathrm{R}_\odot$, a significant amount of deflection and rotation happens between 30~$\mathrm{R}_\odot$ and 1~AU. We compare the flux-rope orientation to the local orientation of the heliospheric current sheet (HCS). We find that slow flux ropes tend to align with the streams of slow solar wind in the inner heliosphere. During the solar-cycle minimum the slow solar wind channel as well as the HCS usually occupy the area in the vicinity of the solar equatorial plane, which in the past led researchers to the hypothesis that flux ropes align with the HCS. Our results show that exceptions from this rule are explained by interaction with the Parker-spiraled background magnetic field, which dominates over the magnetic interaction with the HCS in the inner heliosphere at least during solar-minimum conditions.
\end{abstract}

%%%%%%%%%%%%%%%%%%%%%%%%%%%%%%%%%%%%%%%%%%%%%%%%%%%
%% Keywords
%

\end{opening}
%-------------------------------------------------

%%%%%%%%%%%%%%%%%%%%%%%%%%%%%%%%%%%%%%%%%%%%%%%%%%%
%% Sections
%

\section{Introduction}\label{s:intro} 

Coronal mass ejections (CMEs) are the main drivers of space weather ({\eg} \opencite{Tsurutani1988}; \citeauthor{Huttunen2002b}, \citeyear{Huttunen2002b}; \opencite{Zhang2007}), and thus forecasting their structures at 1~AU is essential. However we do not yet possess detailed knowledge about the way CMEs evolve after their eruption from the Sun. The only mechanism of CME eruption that we know so far is via the ejection of a flux rope from the Sun \cite{Chen2011}. The majority of CMEs are thus believed to have flux ropes at their cores. However, only about 40\,\% of observed CMEs exhibit well-defined flux-rope signatures and can be called flux-rope CMEs (FR-CMEs: \citeauthor{Vourlidas2012}, \citeyear{Vourlidas2012}). We will leave aside the question of whether all CMEs are FR-CMEs and focus on another important aspect of this phenomenon, which is the evolution of flux ropes after their ejection from the Sun. When discussing CMEs we will refer to FR-CMEs.

The flux-rope evolution can be decomposed into expansion, latitudinal and longitudinal deflections, rotation, and distortion, where the set of latitudinal and longitudinal deflections and rotation describes the change of global orientation of a flux rope. Geometrically all of these components are coupled together and are hard to study separately \cite{NievesChinchilla2012}. For instance, neglecting the possible deflections of a flux rope would affect our perception of its rotation or \textit{vice versa}. That means that the only way to reliably study the geometrical evolution of a flux rope is by treating it as a three-dimensional (3D) object from the start. 

The evolution of a flux rope directly affects its geomagnetic effectiveness. Deflections may cause a limb CME to hit the Earth or a halo CME to miss the Earth. Rotation of a flux rope, which is about to hit the Earth can change the efficiency of magnetic reconnection between the flux rope and the magnetosphere of the Earth, thus altering the geoeffectiveness of the impact.

% longitudinal deflection
The CME evolution is associated with its interaction with the magnetic field of the Sun and with the background solar wind and/or surrounding magnetic structures embedded into the solar wind. \inlinecite{Wang2004} suggested that the longitudinal deflection of CMEs may be explained by the interaction with Parker-spiral-structured solar wind. Slow CMEs are pushed by faster solar wind thus feeling a westward component of force while fast CMEs are blocked by slower solar wind thus feeling an eastward component of force. \inlinecite{Gopalswamy2009} suggested that driverless shocks detected at Earth with associated CMEs that originated from the central solar meridian can be explained by the kinematic interaction of CMEs with nearby coronal holes that deflected CMEs away from the Sun--Earth line. Halo CMEs originating from the solar limb can cause geomagnetic storms ({\eg} \citeauthor{Huttunen2002b}, \citeyear{Huttunen2002b}). According to the statistical study by \inlinecite{Gopalswamy2008}, about 9\,\% of large geomagnetic storms are caused by limb halos. \inlinecite{Cid2012} also showed that in particular CMEs originating from the western solar limb can be geoeffective because of longitudinal deflection. \inlinecite{Lugaz2012} demonstrated that a collision with another CME can cause longitudinal deflection. The authors used white-light and {\insitu} observations to make an estimate of longitudinal deflection of the order of $8^\circ$\,--\,$10^\circ$. Another example of collision of two CMEs was studied by \inlinecite{Shen2012}. According to them, the analyzed event resembles the properties of a super-ellastic collision, which caused CMEs' deflection and affected their propagation velocities. \inlinecite{Rodriguez2011} linked coronagraph observations of a CME and {\insitu} measurements of associated magnetic cloud for a number of selected events, and they concluded that, in general, predictions of the flux-rope detections near 1~AU based on the coronagraphic data match the actual {\insitu} observations well. However, in some cases CMEs seemed to have experienced a strong longitudinal deflection in the inner heliosphere.

% latitudinal deflection
It is well-established that many CMEs, in particular near solar minimum, tend to deflect towards the Solar Equator \cite{Plunkett2001,Cremades2006,Wang2011}. \inlinecite{Shen2011} explained the latitudinal deflection of slow CMEs in the lower corona by interaction with the background magnetic field perturbed by the CME itself. The magnetic-field lines, which go around the CME become significantly compressed and the free energy from the compression provides a restoring force that acts on the CME, so that the CME tends to deflect into the region with lower magnetic-energy density. However the density of the heliospheric magnetic field decreases with increasing heliocentric distance, which means that the restoring force is significant only close to the Sun.

% rotation
\inlinecite{Yurchyshyn2009} showed in a statistical study of a hundred CMEs in the lower corona that there is a slight preference in CME rotation toward the solar equatorial plane and heliospheric current sheet (HCS), and they suggested that the rotation of CMEs is due to the presence of a heliospheric magnetic field. The flux-rope tilt was estimated in that work by fitting an ellipse to halo-CME coronagraph images. \inlinecite{Lynch2009} showed in MHD simulations that flux ropes created by flare reconnection undergo significant rotation during their propagation through approximately two~$\mathrm{R}_\odot$. They also suggested that the rotation effects in the lower corona might be completely washed out or dominated by the larger-scale streamer-belt orientation or coronal-hole structure when the CME evolves to heliospheric sizes. There are numerous proofs of the latter; {\eg} \inlinecite{Vourlidas2011} showed an example of a CME which kept rotating in the inner heliosphere, possibly due to interaction with a fast stream. \inlinecite{Yurchyshyn2007}, in a study of 25 CME-interplanetary CME pairs, concluded that about one-third of halo CMEs experience rotation of more than $45^\circ$ during their propagation through interplanetary media from the Sun to the Earth.

Most of the studies described above analyzed CME evolution by tracing some geometrical parameters without consistency of treating it as a 3D object in 3D space. A detailed analysis of the evolution of a CME as a 3D structure was presented by \inlinecite{Isavnin2013a} (hereafter referred to as \href{#cite.Isavnin2013a}{Article~I}). The authors studied the deflections and rotations of 15 slow CMEs observed during the decay of Solar Cycle 23 and rise of Solar Cycle 24. The analysis verified that the flux ropes tend to deflect toward the solar equatorial plane but they also experience rotation on their way to 1~AU. In the current article (Article~II) we extend the technique towards a more precise determination of geometrical parameters of the studied CMEs. We make use of observations of solar sources of CMEs, white-light observations and {\insitu} measurements. In this study we estimate the geometrical evolution of CMEs between $\approx$~30~$\mathrm{R}_\odot$ and 1~AU by propagating them as 3D structures through MHD simulated background solar wind. In \href{#cite.Isavnin2013a}{Article~I} we used only the average values of solar-wind velocity and CME propagation speed at 1~AU. In this article we mostly focus on CME rotations without neglecting latitudinal and longitudinal deflections. Our technique allows us to study CME evolution in relation to the HCS. We seek answers to the following questions:
\begin{itemize}
\item at what heliocentric distances do most of the rotation and deflection of flux ropes occur? and
\item why and how do the flux ropes tend to align with the HCS?
\end{itemize}

\section{Methodology}\label{s:method}

\begin{figure}
\centerline{
\includegraphics[width=0.6\textwidth,clip=]{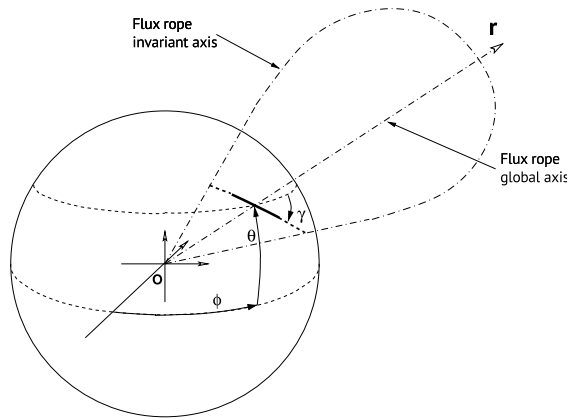}
}
\caption{Scheme of a flux rope ejected from the Sun. The vector $\mathbf{r}$ represents the global axis of the flux rope originating in the Sun and going through the apex of the flux rope. It defines the latitude [$\theta$] and longitude [$\phi$] of the flux rope in Stonyhurst coordinates. The rotation of the flux rope around vector $\mathbf{r}$ (a vector from the Sun to the apex of the flux rope) defines the tilt angle [$\gamma$] of the flux rope. The set of angles $[\theta,\phi,\gamma]$ represents the 3D orientation of the flux rope. (Adapted from Thernisien {\it et al.}, (2009).)}\label{fig:fr_orientation}
\end{figure}

In this section we describe different observational and analytical techniques to trace the flux rope as a 3D structure from its ejection on the Sun to its arrival at 1~AU. The 3D orientation of a flux rope is defined by the latitude [$\theta$] and longitude [$\phi$] of its global axis and the tilt angle [$\gamma$] as shown in Figure \ref{fig:fr_orientation}. We use Stonyhurst coordinates.

We estimate the 3D orientation of a flux rope close to the Sun shortly after the moment that it was ejected by determining the tilt of the associated post-eruption arcades or eruptive prominences (\citeauthor{Yurchyshyn2010}, \citeyear{Yurchyshyn2010} and references therein). The flare model suggested by \inlinecite{Forbes2000} predicts the formation of flare ribbons underneath and parallel to the axis of the erupting flux rope which could be used as proxies of the orientation of the erupted flux rope. Prominence eruptions on the other hand are related to CMEs in a way that part of an erupting filament becomes the bright core of the CME \cite{Gopalswamy2003}. We measure the geometrical orientations of these features using extreme ultraviolet (EUV) observations provided by \textit{EUV Imaging Telescope} (EIT: \citeauthor{Delaboudiniere1995}, \citeyear{Delaboudiniere1995}) and \textit{Extreme UltraViolet Imager}(EUVI: \citeauthor{Wulser2004}, \citeyear{Wulser2004}) instruments onboard \textit{SOlar and Heliospheric Observatory} (SOHO: \citeauthor{Domingo1995}, \citeyear{Domingo1995}) and \textit{Solar TErrestrial RElations Observatory} (STEREO: \citeauthor{Kaiser2008}, \citeyear{Kaiser2008}) spacecraft respectively.

Next, we use coronagraph observations to trace the flux rope from $\approx$~2~$\mathrm{R}_\odot$ to $\approx$~30~$\mathrm{R}_\odot$. We use COR1 ($1.5-4$~$\mathrm{R}_\odot$) and COR2 ($2.5-15$~$\mathrm{R}_\odot$) coronagraphs of the \textit{Sun Earth Connection Coronal and Heliospheric Investigation} (SECCHI) mission onboard STEREO \cite{Howard2008} and C2 ($2.2-6$~$\mathrm{R}_\odot$) and C3 ($3.5-30$~$\mathrm{R}_\odot$) LASCO coronagraphs onboard SOHO \cite{Brueckner1995}. By combining white-light observations of the same flux rope from different view angles we take advantage of 3D forward modeling (FM: \citeauthor{Thernisien2009}, \citeyear{Thernisien2009}) to obtain its geometrical parameters.

After $\approx$~30~$\mathrm{R}_\odot$ we face a problem. Although white-light observations from heliospheric imagers HI-1 and HI-2 are available, a flux rope is not always sufficiently clear to obtain a reliable FM fit.

\begin{figure}
\centerline{
\includegraphics[width=0.7\textwidth,clip=]{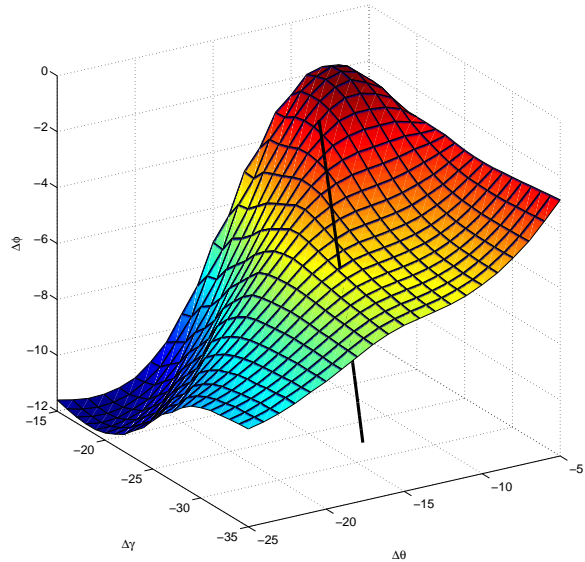}
}
\caption{The ensembles of possible solutions for latitudinal and longitudinal deflections and rotation experienced by a flux rope during its travel from $\approx$~30~$\mathrm{R}_\odot$ to 1~AU depicted in 3D space $[\Delta\theta,\Delta\phi,\Delta\gamma]$. The black curve represents the solutions for the technique presented in Article~I, the surface represents the solutions for the technique presented in Article~II (current paper). The intersection of the curve and the surface gives the unique set of deflections and rotation $[\Delta\theta,\Delta\phi,\Delta\gamma]$ for the studied event. The color scale is used here to improve visual perception of the plot and corresponds to $\Delta\phi$. The presented plot shows the solutions for the event \#5 (27 December 2008) in Table \ref{tbl:results}.}\label{fig:methods_intersect}
\end{figure}

A method for estimating the amount of deflection and rotation that the flux ropes experience between $\approx$~30~$\mathrm{R}_\odot$ to 1~AU was described in \href{#cite.Isavnin2013a}{Article~I}. The method used white-light coronagraph observations, {\insitu} measurements of the associated magnetic cloud at 1~AU, and an estimation of the longitudinal deflection as an input. The longitudinal deflection is calculated using the following equation
\begin{equation}\label{eq:lng_def}
\Delta\phi = \Omega\left(\frac{1}{V_{\mathrm{FR}}}-\frac{1}{V_{\mathrm{SW}}}\right)\mathrm{1AU},
\end{equation}
which describes the kinetics of the interaction of a flux rope with Parker-spiral-structured solar wind \cite{Wang2004}, where $\Delta\phi$ is longitudinal deflection, $\Omega$ is the angular speed of rotation of the Sun, $V_{\mathrm{FR}}$ is the radial velocity of the flux rope leading edge, and $V_{\mathrm{SW}}$ is the solar-wind radial velocity. The underlying assumption is that the magnetic field strength of the FR-CMEs is weak and comparable to the background magnetic field. This seems an appropriate assumption for our events occurring during an unusually weak solar minimum. For the applicability of the technique, it is necessary that the magnetic cloud associated with the flux rope is measured by one of the STEREO spacecraft or \textit{Wind} spacecraft at 1~AU. Grad--Shafranov reconstruction (GSR: \citeauthor{Hu2002}, \citeyear{Hu2002}; \citeauthor{Moestl2009}, \citeyear{Moestl2009}; \citeauthor{Isavnin2011}, \citeyear{Isavnin2011}) was used to estimate the flux-rope cross-section at 1~AU. The most important output of GSR in this case is the local orientation of the invariant axis of the flux rope, which acts as a constraint for the global orientation of the flux rope at 1~AU. Assuming that the flux rope expands self-similarly while preserving the hollow-croissant shape described by FM it is possible to establish a unique trigonometrical relation between the longitudinal deflection and a pair of latitudinal deflection and rotation that the structure experiences on its way from $\approx$~30~$\mathrm{R}_\odot$ to 1~AU:
\begin{equation}\label{eq:old_tech}
(\Delta\theta,\Delta\gamma)=f(\Delta\phi),
\end{equation}
where $\Delta\theta$, $\Delta\phi$, and $\Delta\gamma$ are latitudinal deflection, longitudinal deflection and rotation experienced by a flux rope on the way from $\approx$~30~$\mathrm{R}_\odot$ to 1~AU respectively. The details of the trigonometrical relation (\ref{eq:old_tech}) are described in \href{#cite.Isavnin2013a}{Article~I}. One can think of Equation (\ref{eq:old_tech}) as a curve in 3D space $[\Delta\theta,\Delta\phi,\Delta\gamma]$ (see Figure \ref{fig:methods_intersect}).

A serious limitation of the technique as presented in \href{#cite.Isavnin2013a}{Article~I} is the assumption that the flux-rope leading-edge speed [$V_{\mathrm{FR}}$] and solar-wind radial velocity [$V_{\mathrm{SW}}$] are constant between $\approx$~30~$\mathrm{R}_\odot$ and 1~AU. Obviously, the speed of the leading edge and the solar-wind speed can change during propagation. This problem can be solved by converting Equation (\ref{eq:lng_def}) into integral form:
\begin{equation}\label{eq:lng_def_int}
\Delta\phi = \Omega\int_{r_0}^{1\mathrm{AU}}\left(\frac{1}{V_{\mathrm{FR}}(r)}-\frac{1}{V_{\mathrm{SW}}(r)}\right)\mathrm{d}r,
\end{equation}
where $r_0$ is the height at which we start to trace the flux rope; in fact $r_0$ is the last height where clear white-light observations of the structure from both STEREO spacecraft and SOHO are available. $V_{\mathrm{FR}}(r)$ and $V_{\mathrm{SW}}(r)$ are radial-velocity profiles of the flux rope leading edge and solar wind along the trajectory of the flux rope respectively. Equation (\ref{eq:lng_def_int}) can be solved numerically if the velocity profiles are known.

\inlinecite{Poomvises2010} showed that the velocity of the flux-rope leading edge converges quickly in interplanetary space to match the speed of the surrounding solar wind, so that it becomes nearly constant after 50~$\mathrm{R}_\odot$. This result means that CME drag happens during an early stage of CME evolution and implies that the constant speed of the flux rope leading edge after 50~$\mathrm{R}_\odot$ is a reasonable assumption \cite{Colaninno2013}. In addition, \inlinecite{Kilpua2012} studied Sun-to-Earth travel times of CMEs during the Solar Cycle 23/24 minimum and found that the ICME leading-edge speed measured at 1~AU yielded the best correlation between the predicted and observed travel times. These results justify the usage of the plasma bulk velocity [$V_\mathrm{FR}$] measured \textit{in situ} for the associated magnetic cloud at 1~AU as an estimate for the leading-edge speed. Although HI observations are usually too faint for applying FM, the flux-rope leading edge can be observed quite clearly in them using time-elongation maps (J-maps). Using the harmonic-mean approach (HM: \citeauthor{Lugaz2009}, \citeyear{Lugaz2009}) it is possible to estimate the average speed of the leading edge of the flux rope and its propagation direction by assuming that the leading edge of the flux rope is spherical. Whenever the geometrical configuration of spacecraft makes it possible we apply HM approach with fixed propagation direction obtained with FM at $\approx$~30~$\mathrm{R}_\odot$ and compare the estimated leading edge speed with the {\insitu} speed at 1~AU. Thus we ensure that the use of the leading-edge speed measured \textit{in situ} at 1~AU as an estimate for the speed of its propagation from $\approx$~30~$\mathrm{R}_\odot$ to 1~AU is a reasonable assumption.

We use the results of global 3D MHD simulations carried out with the MAS (Magnetohydrodynamics Around a Sphere) model \cite{Mikic1999,Linker1999,Riley2007,Riley2011} and provided by the Predictive Science STEREO modeling support project (\url{imhd.net/stereo}) as an estimation for the radial velocity of the solar wind [$V_{\mathrm{SW}}(r)$]. The MAS model uses a photospheric magnetic-field map built up from a sequence of observations centered at central meridian over a 27-day period as an input. The resulting global simulation does not include the expanding flux rope and thus the resulting solar wind is not yet perturbed and can be considered as background solar wind. The model is calculated once per Carrington rotation. Thus, it does not take into account solar-wind streams' evolution during one Carrington rotation. This is not so crucial during the solar-minimum period covered by our observations. The use of outdated photospheric magnetic field information would make the results of MHD simulations less reliable during solar maximum when the configuration of magnetic field is subject to frequent and rapid changes. The photospheric magnetic-field map measured during one full Carrington rotation uses the newest data for the western heliosphere and the oldest for the eastern heliosphere which makes the MHD-simulated background solar wind less reliable for the eastern heliosphere and introduces uncertainties in the analysis.

\begin{figure}
\centerline{
\includegraphics[width=0.3\textwidth,clip=]{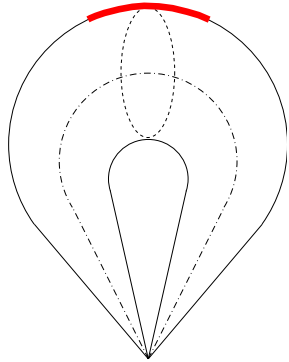}
}
\caption{The scheme of the flux rope. The thick red curve shows the part of the leading edge that we are tracing through the background solar wind. The radial velocity of the background solar wind is averaged along the leading edge of the flux rope.}\label{fig:fr_leading_edge}
\end{figure}

Now that the flux-rope expansion velocity and background solar-wind velocity are estimated, we can solve Equation (\ref{eq:lng_def_int}) numerically. Equation (\ref{eq:lng_def_int}) describes only longitudinal deflection of the flux rope. We solve it for the range of possible latitudinal deflection and rotation angles [$\Delta\theta$ and $\Delta\gamma$], respectively. For simplicity we assume that $\theta$ and $\gamma$ change linearly from $r_0$ to 1~AU:
\begin{equation}\label{eq:theta_linear}
\theta(r)=\theta(r_0)+\Delta\theta\f{r-r_0}{1\mathrm{AU}-r_0},
\end{equation}
\begin{equation}\label{eq:gamma_linear}
\gamma(r)=\gamma(r_0)+\Delta\gamma\f{r-r_0}{1\mathrm{AU}-r_0}.
\end{equation}
We do not assume a point object for the FR tracing. Instead we trace the front part of its leading edge as shown in Figure \ref{fig:fr_leading_edge}. In each integration step [$n$] we calculate the longitude [$\phi_n$]  of the flux rope with the following equation:
\begin{equation}\label{eq:lng_step}
\phi_n=\phi_{n-1}+\Omega\left(\f{1}{V_{\mathrm{FR}}(r_n)}-\f{1}{\tilde{V}_{\mathrm{SW}}(r_n)}\right)\left(r_n-r_{n-1}\right),
\end{equation}
where $\tilde{V}_{\mathrm{SW}}(r_n)$ is the background solar-wind radial velocity averaged along the part of the leading edge of the flux rope as depicted in Figure \ref{fig:fr_leading_edge}. The background solar wind simulated with the MAS model for one Carrington rotation is static, {\ie} no dynamic outflow is simulated. If we assume that the photospheric magnetic field has a stationary configuration then the steady outflow of the solar wind can be modeled by rotating the MHD-simulated solar wind as a 3D object with the solar angular-rotation speed [$\Omega$]. The rotation is taken into account in each integration step.

The described technique can be represented with the following equation:
\begin{equation}\label{eq:new_tech}
\Delta\phi=g(\Delta\theta,\Delta\gamma).
\end{equation}
Equation (\ref{eq:new_tech}) describes also a surface in 3D space $[\Delta\theta,\Delta\phi,\Delta\gamma]$ (see Figure \ref{fig:methods_intersect}).

Finally, we combine the two techniques. The intersection of the curve described by Equation (\ref{eq:old_tech}) and the surface described by Equation (\ref{eq:new_tech}) define the unique set of longitudinal and latitudinal deflections and rotation of the flux rope that is the solution for both methods simultaneously as shown in Figure~\ref{fig:methods_intersect}.

To recap, we use the following methodology to trace ejected flux ropes from the Sun to 1~AU. First, we use extreme ultraviolet observations of flux-rope signatures close to the solar surface to estimate the initial orientation of the FR. Then, we use multipoint white-light observations to model the flux rope with FM from $\approx$~2~$\mathrm{R}_\odot$ to $\approx$~30~$\mathrm{R}_\odot$. Finally, we trace the flux rope through the MHD-simulated solar wind from $\approx$~30~$\mathrm{R}_\odot$ to 1~AU using {\insitu} measurements of the associated magnetic cloud as a local constraint for the global orientation of the flux rope.

\section{Results}

We examine the 14 events during the decay of Solar Cycle 23 and rise of Solar Cycle 24 presented in \href{#cite.Isavnin2013a}{Article~I} with this technique. We deduce 3D orientation of flux ropes in several points during their propagation. Therefore, unlike \href{#cite.Isavnin2013a}{Article~I}, we always have an estimate of the direction of rotation of flux ropes. The results are presented in Table \ref{tbl:results}. The orientation (latitude and longitude) is given in Stonyhurst coordinates (see Figure \ref{fig:fr_orientation}). The EUVI/EIT part of Table \ref{tbl:results} represents the EUV observations on the solar disk and estimations of the flux-rope orientation using post-eruption arcades or eruptive prominences. The longitudinal expansion of these structures is often unclear, so we estimate only longitude and tilt angle from those observations. The COR1/C2 part of the Table \ref{tbl:results} corresponds to the FM results of white-light images of flux ropes in the field of view of COR1 and C2 coronagraphs. The values of latitude, longitude, and tilt angle are averages of the FM parameters. The average height of the center of the leading edge of the flux rope was in the range of 3 to 8~$\mathrm{R}_\odot$ for those observations. Similarly, the COR2/C3 part of the Table \ref{tbl:results} corresponds to the average 3D orientation of flux ropes in the field of view of COR2 and C3 coronagraphs with average height of the center of the leading edge of 10 to 22~$\mathrm{R}_\odot$. The 1~AU part of the Table \ref{tbl:results} corresponds to the results of the flux-rope propagation technique which was described in Section 2. 
%Event \#10 (23 May 2010) is marked with an asterisk since its evolution was influenced by a collision with another CME. For this event we use the value of longitudinal deflection of $9^\circ$ estimated by \inlinecite{Lugaz2012}. 
The total latitudinal deflections experienced by flux ropes from the Sun to 1~AU ranged from $20^\circ$ to $49^\circ$, the total longitudinal deflections ranged from $-28^\circ$ to $14^\circ$, and the total rotations ranged from $4^\circ$ to $164^\circ$.

\begin{figure}
\centerline{
\includegraphics[width=0.6\textwidth,clip=]{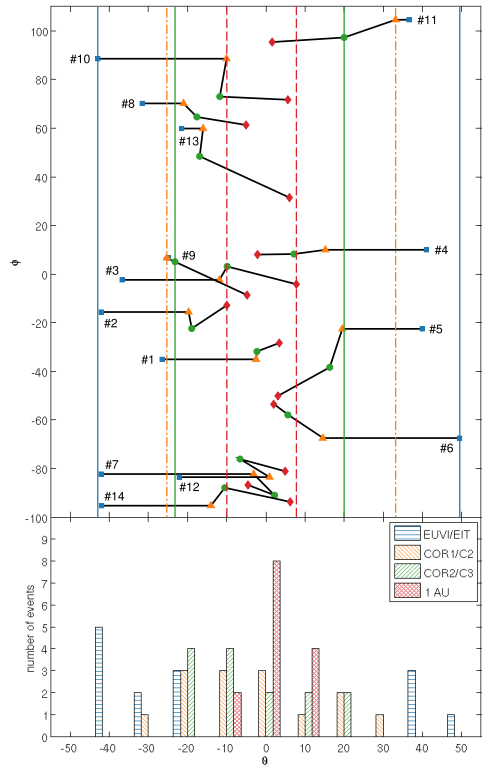}
}
\caption{Latitudinal and longitudinal deflections of 14 analyzed flux ropes (top) and distributions of latitudinal deflections for different heights (bottom). The markers represent the direction of the vector from the Sun to the apex of the flux rope in Stonyhurst coordinates. Blue squares correspond to 1\,--\,2~$\mathrm{R}_\odot$, orange triangles correspond to 2\,--\,8~$\mathrm{R}_\odot$, green circles correspond to 8\,--\,30~$\mathrm{R}_\odot$, red diamonds correspond to 1~AU. Blue, orange, red, and green lines show latitudinal ranges for flux rope orientation at respective heliocentric heights. Note that even though a substantial part of flux-rope deflection happens in the lower corona, the flux ropes' orientation continues to evolve in the inner heliosphere.}\label{fig:deflections}
\end{figure}

\begin{figure}
\centerline{
\includegraphics[width=0.9\textwidth,clip=]{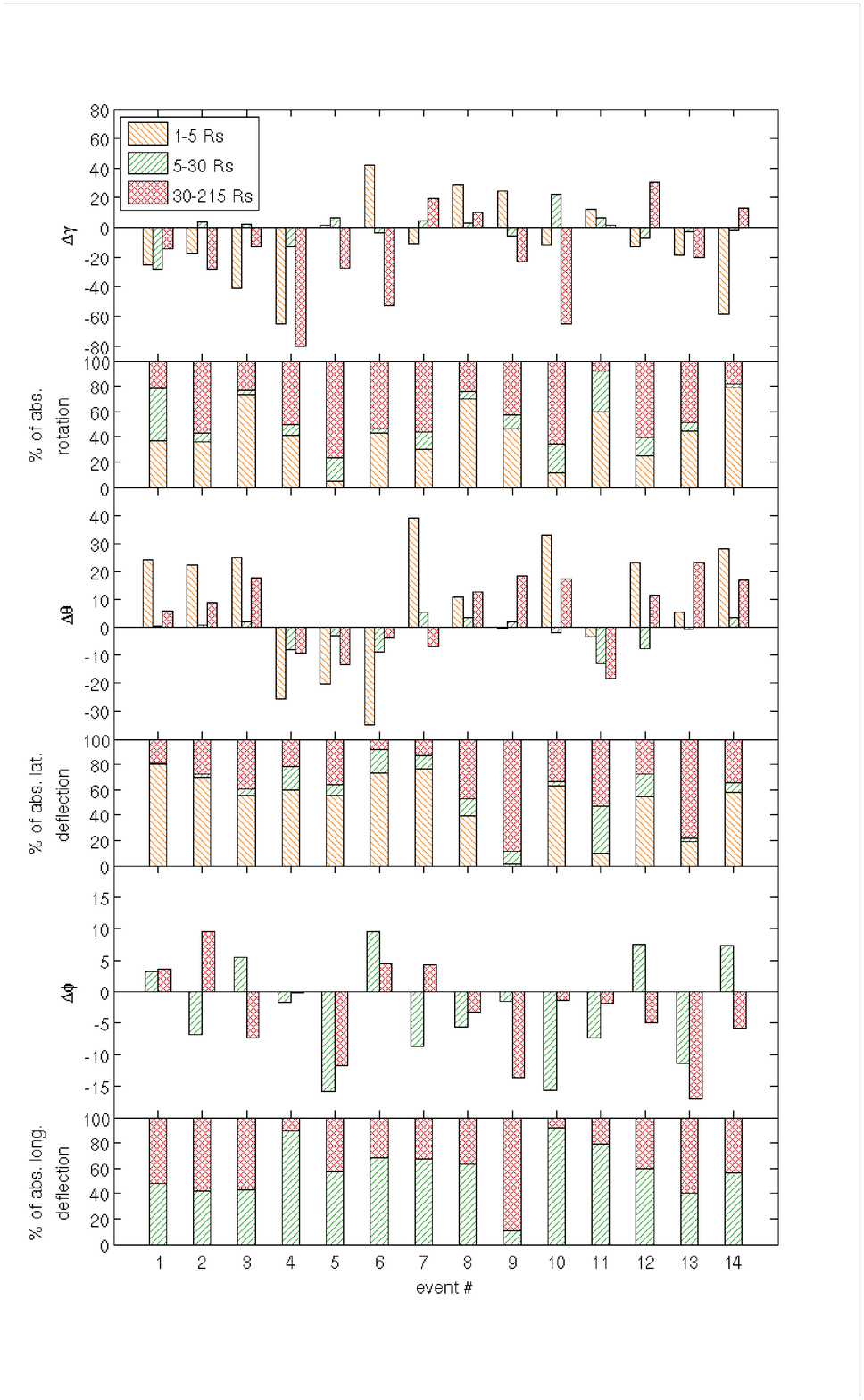}
}
\caption{The evolution of orientations of the 14 studied flux ropes during the following segments of their travel from the Sun to 1 AU: 1\,--\,5~$\mathrm{R}_\odot$, 5\,--\,30~$\mathrm{R}_\odot$, 30~$\mathrm{R}_\odot$\,--\,1~AU. Panels 1, 3, and 5 show the change of tilt angle, latitude, and longitude respectively. Panels 2, 4, and 6 show what part of the overall evolution a flux rope experienced during each segment.}\label{fig:all_stat}
\end{figure}

Figures \ref{fig:deflections} and \ref{fig:all_stat} display the results given in Table \ref{tbl:results}. Figure \ref{fig:deflections} shows the longitudinal and latitudinal deflections of the 14 flux ropes. The markers on the plot correspond to the orientation of the vector from the Sun to the apex of the flux rope. Figure \ref{fig:deflections} confirms that the flux ropes tend to deflect towards the solar equatorial plane. Close to the Sun, the flux ropes fall within the range of latitudes $[-42^\circ;50^\circ]$, at $\approx$~5~$\mathrm{R}_\odot$ the range shrinks to $[-25^\circ;33^\circ]$, at $\approx$~18~$\mathrm{R}_\odot$ it shrinks further to $[-22^\circ;20^\circ]$, and finally at 1~AU the flux ropes fall within the range of latitudes of $[-10^\circ;8^\circ]$. The longitudinal deflections between $\approx$~5~$\mathrm{R}_\odot$ and 1~AU ranged from $-29^\circ$ to $14^\circ$. For some of the events the direction of longitudinal deflection changed on the way to 1~AU ({\eg} \#2, \#3, \#7). Such behavior can be caused by structural variability of the solar wind, {\eg} a flux rope moved in the fast solar wind and then entered an area with slower solar wind. 
%It can be also caused by interaction with other magnetic structures as it may have happened for the event \#10. 
These deviations could be also due to misidentification of source regions, erroneous FM fits and uncertainties of GSR.

To investigate the amount of rotation and deflection at different heliospheric distances we divide the Sun--Earth distance into three segments. The segments chosen are the following: $1-5$~$\mathrm{R}_\odot$ (from EUVI/EIT to COR1/C2 observations), $5-30$~$\mathrm{R}_\odot$ (from COR1/C2 to COR2/C3 observations) and $30-215$~$\mathrm{R}_\odot$ (from COR2/C3 to {\insitu} observations). Figure \ref{fig:all_stat} shows that flux-rope geometrical evolution happens fast in the lower corona, {\ie} $\approx$~41\,\% of estimated flux-rope rotation and $\approx$~48\,\% of latitudinal deflection happens during the first 3\,\% ($\approx$~5~$\mathrm{R}_\odot$) of the path from the Sun to 1~AU. During the first $\approx$~30~$\mathrm{R}_\odot$ of the travel, the flux ropes experience, on average, $\approx$~57\,\% of total rotation, $\approx$~62\,\% of total latitudinal deflection and $\approx$~58\,\% of total longitudinal deflection. However, it is important to note that a significant amount of deflection and rotation happens in the inner heliosphere from 30~$\mathrm{R}_\odot$ to 1~AU. The flux-rope evolution after $\approx$~30~$\mathrm{R}_\odot$ could be caused by several effects: the flux ropes may change their orientation because the initial momentum gained due to magnetic interactions in the lower corona is sufficient to influence the flux-rope evolution in the inner heliosphere; the decreasing magnetic interaction with the solar magnetic field could become dominated by the kinematic interaction with solar streams in the inner heliosphere; the MHD models may not be sufficiently accurate. 

\begin{figure}
\centerline{
\includegraphics[width=0.45\textwidth,clip=]{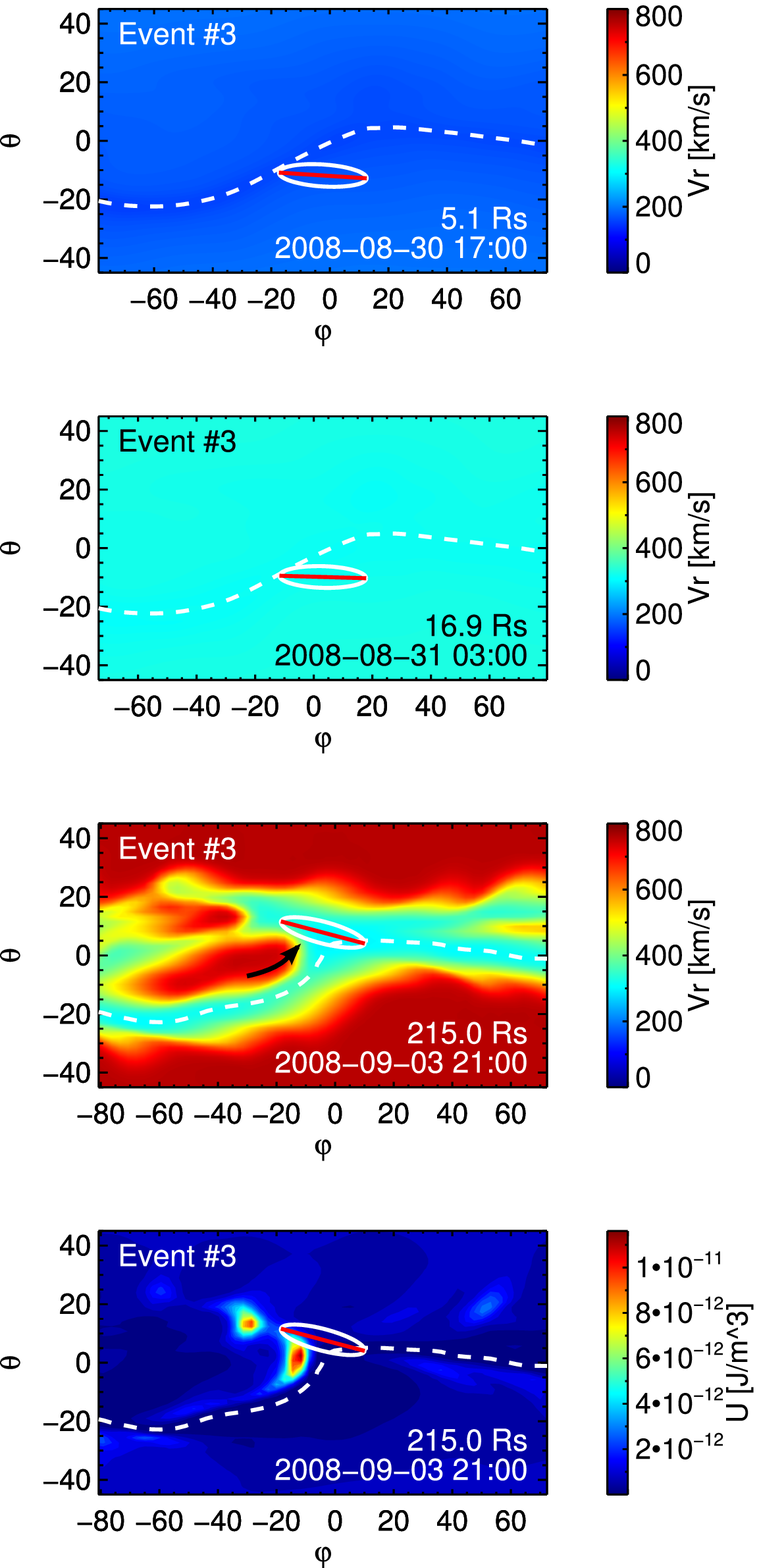}
\includegraphics[width=0.465\textwidth,clip=]{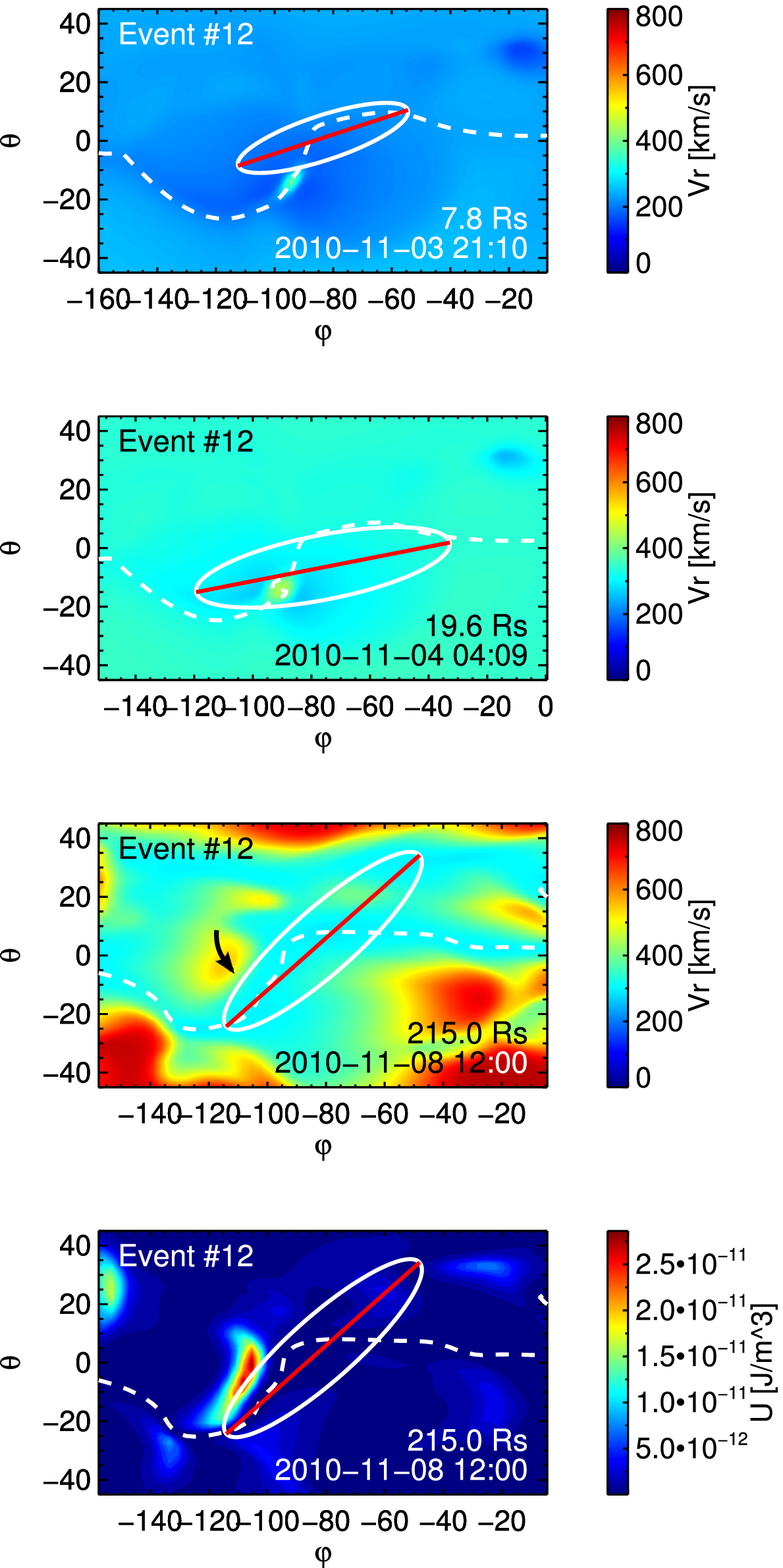}
}
\caption{Maps of the radial component of the background solar wind (top three panels, color-coded), mangetic-energy density (bottom panel, color-coded), and HCS (white dashed curve). The orientation of a flux rope is represented by an ellipse. The size of an ellipse corresponds to the angular size of a flux rope. The black arrows show the approximate direction of momentum acting on flux ropes due to magnetic energy density gradient. Upper three panels correspond to the heights of 2\,--\,8~$\mathrm{R}_\odot$, 8\,--\,30~$\mathrm{R}_\odot$ and 1~AU, respectively. Bottom panels show the magnetic energy density at 1~AU. Events \#3 (left) and \#7 (right) are presented.}\label{fig:hcs_ex1}
\end{figure}

\begin{figure}
\centerline{
\includegraphics[width=0.45\textwidth,clip=]{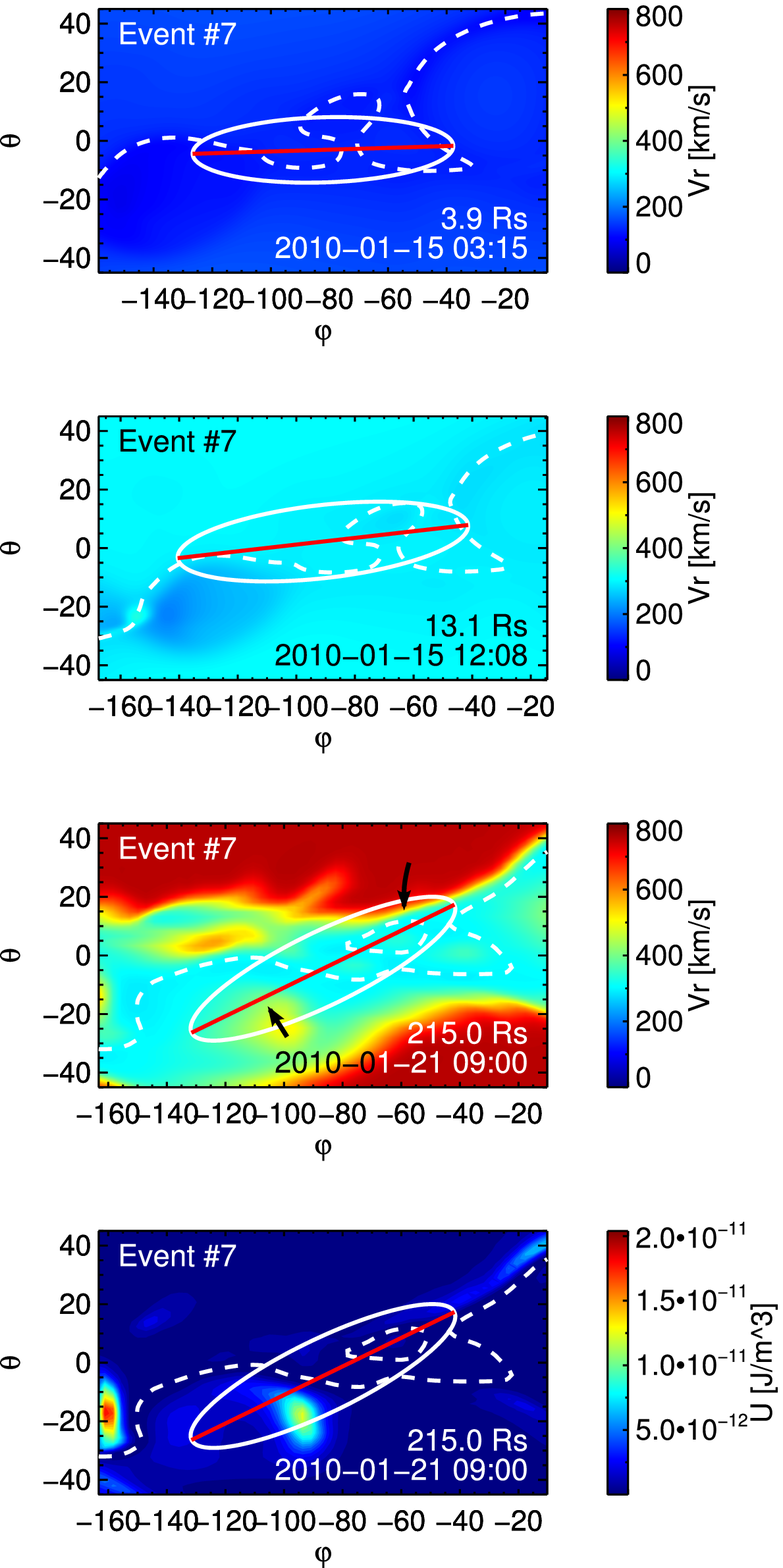}
\includegraphics[width=0.45\textwidth,clip=]{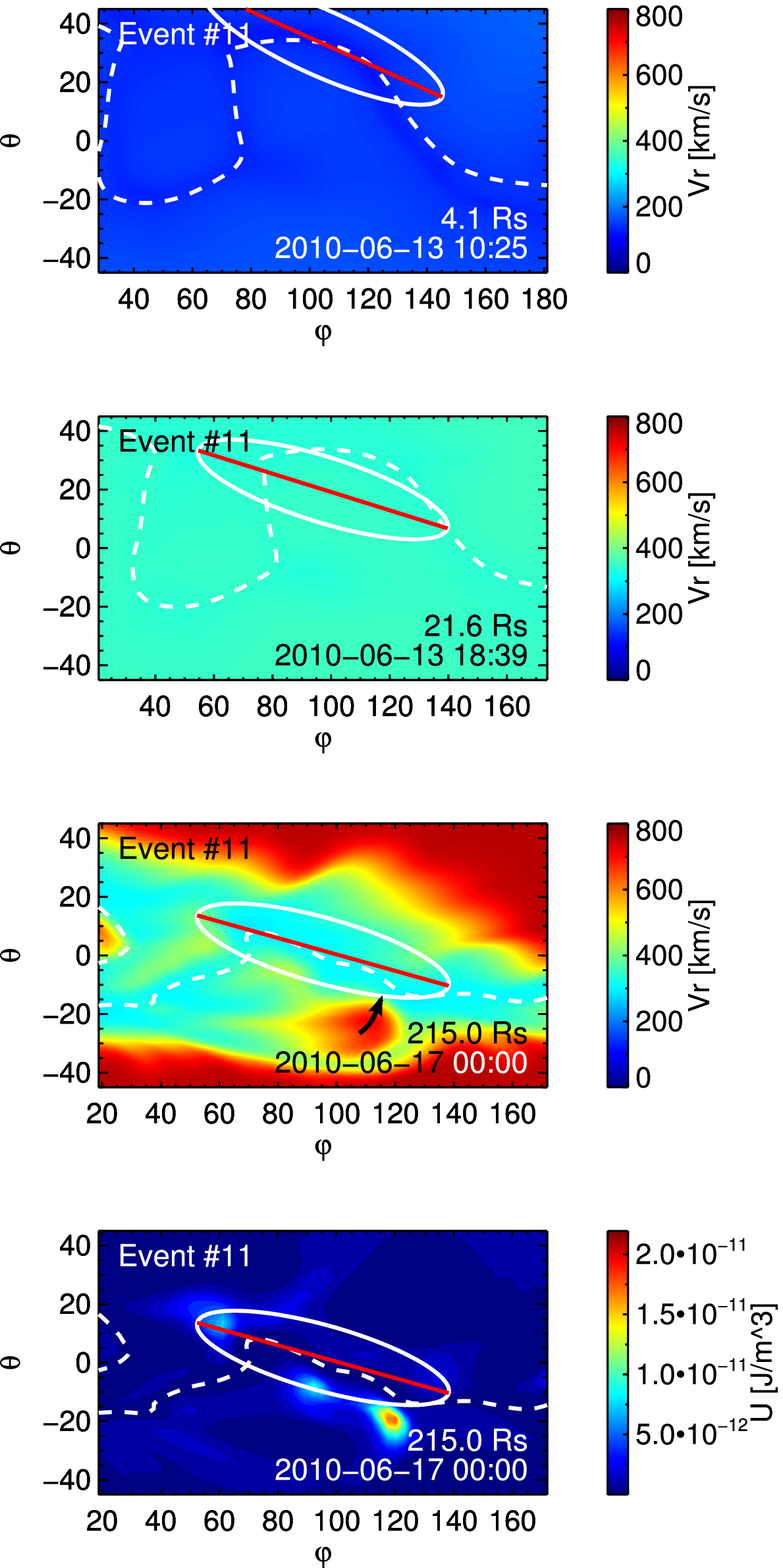}
}
\caption{Maps of the radial component of the background solar wind (top three panels, color-coded), mangetic-energy density (bottom panel, color-coded), and HCS (white dashed curve). The orientation of a flux rope is represented by an ellipse. The size of an ellipse corresponds to the angular size of a flux rope. The black arrows show the approximate direction of momentum acting on flux ropes due to magnetic energy density gradient. Upper three panels correspond to the heights of 2\,--\,8~$\mathrm{R}_\odot$, 8\,--\,30~$\mathrm{R}_\odot$ and 1~AU, respectively. Bottom panels show the magnetic energy density at 1~AU. Events \#11 (left) and \#12 (right) are presented.}\label{fig:hcs_ex3}
\end{figure}

We use constant heliocentric distance slices of the 3D MAS MHD model to study the rotation of flux ropes relative to the local orientation of HCS. Figures \ref{fig:hcs_ex1} and \ref{fig:hcs_ex3} show six events from our study. Each example is presented with three panels corresponding to the heights of the flux-rope leading edge at 2\,--\,8~$\mathrm{R}_\odot$, 8\,--\,30~$\mathrm{R}_\odot$, and 1~AU, respectively. The radial component of the background solar-wind velocity is color-coded. The local HCS orientation is depicted with the white dashed curve, which is estimated as the boundary of the area where the radial component of the background magnetic field is zero. The orientations of flux ropes are shown with white ellipses. The position and tilt of each ellipse depict the geometrical orientation of a flux rope. The size of an ellipse gives an idea about the angular width of a flux rope. It can be seen that the local shape of the HCS is too complex to allow an estimate of the flux rope tilt relative to the HCS with a single value. So we analyze each event by assessing individually images similar to Figures \ref{fig:hcs_ex1} and \ref{fig:hcs_ex3}.

What can be noticed from Figures \ref{fig:hcs_ex1} and \ref{fig:hcs_ex3} is that the flux ropes tend to stay close to the HCS until 1~AU although they are not usually strictly aligned with the local HCS. The radial velocity of the background solar wind becomes more structured with increasing heliocentric distance thus creating well-defined slow and fast solar-wind channels. A wide channel of slow wind is typically formed in the vicinity of the HCS, and our flux ropes (which are relatively slow) expand inside this channel. The orientation of flux ropes inside the slow solar-wind channels seems to be defined by the combination of the channel boundaries itself and those of faster streams inside the channel. Depending on the geometry of the slow solar-wind channel and faster streams inside it the resulting flux-rope configurations can be classified into three groups: flux ropes that are strongly affected by faster streams, and thus not aligned with the HCS (Figure \ref{fig:hcs_ex1}, left); flux ropes that are influenced by faster streams but stay close to the HCS (Figure \ref{fig:hcs_ex3}, right); flux ropes that stay aligned with the HCS and are not affected strongly enough by faster streams to change their orientation (Figure \ref{fig:hcs_ex1}, right; Figure \ref{fig:hcs_ex3}, left). 

Event \#3 (Figure \ref{fig:hcs_ex1}, left) belongs to the first group of events in our classification. It seems that the fast solar-wind stream isolates or compresses a part of the slow solar wind channel with the flux trapped in it. The magnetic energy density map (Figure \ref{fig:hcs_ex1}, left, bottom panel) shows magnetic compression in the region between the fast stream and the flux rope, which could be a signature of their interaction. Note also that the flux rope appears to have crossed over the HCS at 1~AU, which implies that magnetic interactions with the HCS are weak far from the Sun.

Event \#12 (Figure \ref{fig:hcs_ex1}, right) demonstrate the flux rope with orientation at 1~AU influenced by the structured background solar wind. Fast solar wind stream could a possible reason for the origin of compression region between the stream and the flux rope (Figure \ref{fig:hcs_ex1}, right, bottom panel). The flux rope stays close to the HCS but it is not completely aligned with it due to interaction with the faster stream.

The flux ropes that correspond to Event \#7 (Figure \ref{fig:hcs_ex3}, left) and Event \#11 (Figure \ref{fig:hcs_ex3}, right) seem to align well with the HCS at 1~AU. This can lead to an assumption that these flux ropes were attracted to the HCS due to magnetic interaction. However, the hypothesis about interaction with background magnetic field explains the positioning of these flux ropes as well, {\ie} the flux ropes were pushed into the zone of lower magnetic-energy density.

Closer to the Sun ($\leq30$~$\mathrm{R}_\odot$) the radial velocity of background solar wind is more uniform and we do not see any obvious interactions with background solar wind in the first two panels of Figures \ref{fig:hcs_ex1} and \ref{fig:hcs_ex3}. However, flux ropes tend to stay close to the HCS. This behaviour supports the suggestion that close to the Sun the magnetic interactions influence the orientation a flux rope. However, as the flux rope propagates further from the Sun the disturbance of the surrounding magnetic field created by it gets weaker and eventually the interactions with solar-wind streams become the dominant factor influencing the orientation and evolution of the flux rope. The fact that our flux ropes were located close to the HCS in the first place points to either an origin close to the HCS or to strong deflection towards the HCS in the low corona.

\section{Discussion}

In this article we describe a technique for studying the 3D evolution of ejected flux ropes from the Sun to 1~AU. The method uses EUV and white-light observations from SOHO and STEREO and {\insitu} measurements at 1~AU from \textit{Wind} and STEREO. The results of MHD simulations made with the MAS model provided an estimate of the background solar wind when tracing flux ropes from $\approx$~30~$\mathrm{R}_\odot$ to 1~AU. GSR is used to estimate the local orientations of flux ropes at 1~AU which in turn is used as a constraint for their global orientations. An example of FM analysis of white-light observations and GSR analysis of {\insitu} measurements for the Event \#12 was presented in \href{#cite.Isavnin2013a}{Article~I} (Figures 3 and 4).

Using this method we studied 14 ejected flux ropes observed during the decay of Solar Cycle 23 and early rise of Solar Cycle 24. Our analysis confirmed that flux ropes tend to deflect towards the solar equatorial plane, and we quantified the amount of deflection at different heliocentric distances. We find that a large part of the latitudinal deflection occurs within a few $\mathrm{R}_\odot$. Thus, magnetic interactions with nearby coronal holes may be the main cause of latitudinal deflection \cite{Shen2011}. Our results show that about 60\,\% of the geometrical evolution of flux ropes happens within the first 30~$\mathrm{R}_\odot$. Nevertheless, a significant part of the deflection and rotation occurs between 30~$\mathrm{R}_\odot$ and 1~AU, {\ie} outside the current coronagraph fields of view.

We also study the rotation of the flux ropes relative to the HCS. As expected, we find that the flux ropes stay close the HCS but they are not necessarily aligned with it. Some non-alignments seem to be influenced by the relative location of fast and slow-wind streams ({\eg} Figure \ref{fig:hcs_ex1}(right)) while others may be due to modeling errors. However, the latter should be minimal as our analysis extends over one of the deepest solar minima on record.

We studied the evolution of slow solar flux ropes during the low solar activity phase in \href{#cite.Isavnin2013a}{Article~I} and here to develop a technique for tracing flux ropes from the Sun to 1~AU. We expect that faster flux ropes, observed more frequently during solar-maximum conditions, should experience higher degrees of distortion and longitudinal eastward deflection when propagating through slower background solar wind. The differential background solar wind can also be responsible for distortion of flux ropes. \inlinecite{Savani2010} studied a CME that was distorted into a concave structure apparently due to kinematic interaction with slow solar wind. Such distortion should be more pronounced for fast CMEs since the drag on the CME is stronger. If the background solar wind is highly structured the drag force will be different for different parts of the flux ropes, causing distortion.

\begin{landscape}
\begin{table}[!ht]
\begin{tabular}{r || l || r | r || r | r | r | r || r | r | r | r || r | r | r | r}
\multirow{2}{*}{\#} & CME & \multicolumn{2}{l}{0\,--\,2~$\mathrm{R}_\odot$~(EUVI/EIT)} & \multicolumn{4}{l}{2\,--\,8~$\mathrm{R}_\odot$~(COR1/C2)} & \multicolumn{4}{l}{8\,--\,30~$\mathrm{R}_\odot$~(COR2/C3)} & \multicolumn{4}{l}{1~AU} \\
& Date & $\theta$ & $\gamma$ & $\theta$ & $\phi$ & $\gamma$ & $h_{\mathrm{LE}}$ & $\theta$ & $\phi$ & $\gamma$ & $h_{LE}$ & $\theta$ & $\phi$ & $\gamma$ & $V_{\mathrm{FR}}$ \\
\hline
1 & 02 June 2008 & -27 & 5 & -3 & -35 & -20 & 3.9 & -2 & -32 & -49 & 17.0 & 4 & -28 & -63 & 397 \\
2 & 07 July 2008 & -42 & 12 & -20 & -16 & -6 & 4.2 & -19 & -22 & -2 & 18.5 & -10 & -13 & -30 & 561 \\
3 & 30 August 2008 & -37 & 38 & -12 & -2 & -4 & 5.1 & -10 & 3 & -2 & 16.9 & 8 & -4 & -15 & 466 \\
4 & 12 December 2008 & 41 & 141 & 15 & 10 & 76 & 4.6 & 7 & 8 & 63 & 20.6 & -2 & 8 & -17 & 355 \\
5 & 27 December 2008 & 40 & 5 & 20 & -23 & 7 & 3.6 & 16 & -38 & 14 & 14.0 & 3 & -50 & -14 & 425 \\
6 & 27 September 2009 & 50 & -35 & 15 & -67 & 7 & 3.5 & 6 & -58 & 4 & 10.8 & 2 & -54 & -49 & 362 \\
7 & 15 January 2010 & -42 & 12 & -3 & -82 & 2 & 3.9 & 2 & -91 & 7 & 13.1 & -5 & -87 & 26 & 330 \\
8 & 01 February 2010 & -32 & -7 & -21 & 70 & 22 & 5.6 & -18 & 65 & 25 & 15.6 & -5 & 61 & 35 & 402 \\
9 & 03 April 2010 & -25 & -10 & -25 & 7 & 15 & 3.2 & -23 & 5 & 9 & 20.9 & -5 & -9 & -14 & 750 \\
%10* & 2010-05-23 & 19 & -45 & 16 & 0 & -17 & 4.4 & 6 & 12 & 55 & 14.6 & -8 & 3 & 118 & 354 \\
10 & 27 May 2010 & -43 & 51 & -10 & 89 & 40 & 5.8 & -12 & 73 & 62 & 16.4 & 6 & 72 & -3 & 408 \\
11 & 13 June 2010 & 37 & -37 & 33 & 105 & -24 & 4.1 & 20 & 97 & -17 & 21.6 & 2 & 95 & -16 & 500 \\
12 & 03 November 2010 & -22 & 31 & 1 & -84 & 18 & 7.9 & -7 & -76 & 11 & 19.5 & 5 & -81 & 42 & 399 \\
13 & 12 December 2010 & -22 & 28 & -16 & 60 & 9 & 4.4 & -17 & 48 & 6 & 13.9 & 6 & 31 & -14 & 482 \\
14 & 12 December 2010 & -42 & 50 & -14 & -95 & -9 & 4.7 & -11 & -88 & -11 & 20.5 & 6 & -94 & 2 & 350 \\
\end{tabular}
\caption{Analysis of the evolution of ejected flux ropes for 14 events. Latitude [$\theta$], longitude [$\phi$], and tilt [$\gamma$] represent the geometrical orientation of the flux rope at different stages of its evolution. $h_{LE}$ is the height of the leading edge of the flux ropes above the Sun measured in $\mathrm{R}_\odot$. $V_{\mathrm{FR}}$ is the flux-rope leading-edge speed measured {\insitu} at 1~AU. 
%Event \#10 marked with asterisk is the case of the flux rope which experienced collision with another flux rope and thus its evolution was affected by that collision.
}
\label{tbl:results}
\end{table}
\end{landscape}

%%%%%%%%%%%%%%%%%%%%%%%%%%%%%%%%%%%%%%%%%%%%%%%%%%%%%%%%%%%%%%%%%%%%%%%%%%%
%% Appendix
%
% \appendix   

%%%%%%%%%%%%%%%%%%%%%%%%%%%%%%%%%%%%%%%%%%%%%%%%%%%%%%%%%%%%%%%%%%%%%%%%%%%
%% Acknowledgements
%
\begin{acks}
The work of AI and EK was supported by the Academy of Finland. The work of AV is supported by NASA contract S-136361-Y to the Naval Research Laboratory. LASCO was constructed by a consortium of institutions: NRL (USA), MPI f{\"u}r Aeronomie (Germany), LAS (France), and University of Birmingham (UK). The SECCHI data are produced by an international consortium of the NRL, LMSAL, and NASA GSFC (USA), RAL and University of Birmingham (UK), MPS (Germany), CSL (Belgium), IOTA, and IAS (France).
\end{acks}

%%% %%%%%%%%%%%%%%%%%%%%%%%%%%%%%%%%%%%%%%%%%%%%%%%%%%%%%%%%%%%
%% Bibliography
%
% Using BibTeX
%
\bibliographystyle{spr-mp-sola}
% %\bibliographystyle{spr-mp-sola-cnd} %% Alternative style: no title, no concluding page
\bibliography{bibliography.bib}  
%
% Without BibTeX 
% \begin{thebibliography}{}
% \bibitem[\protect\citeauthoryear{Author}{Year}]{key}
%   <bibliographical entry>
%
% \bibitem[\protect\citeauthoryear{}{}]{}
%   
%  
% \end{thebibliography}

\end{article} 
\end{document}